\documentstyle[prl,aps,epsf,twocolumn]{revtex}

\begin{document}
\draft

\title{Formation of a rotating hole from a close limit
head-on collision}

\author{William Krivan}
\address{Department of Physics, University of Utah,
         Salt Lake City, UT 84112 \\
    and Institut f\"ur Astronomie und Astrophysik, Universit\"at 
        T\"ubingen, D-72076 T\"ubingen, Germany}
\author{Richard H. Price}
\address{
Department of Physics, University of Utah, Salt Lake City, UT 84112}
\maketitle
\begin{abstract}
Realistic black hole collisions result in a rapidly rotating Kerr
hole, but simulations to date have focused on nonrotating final
holes. Using a new solution of the Einstein initial value equations we
present here waveforms and radiation for an axisymmetric
Kerr-hole-forming collision starting from small initial separation
(the ``close limit'' approximation) of two identical rotating
holes. Several new features are present in the results: (i) In the limit
of small separation, the waveform is linear (not quadratic) in the separation.
(ii) The waveforms show damped oscillations mixing quasinormal
ringing of different multipoles.
\end{abstract}


The study of black hole collisions with numerical relativity
\cite{grandchallenge} will give unprecedented insights into the
nonlinear workings of relativistic gravitation in strong field
situations and is expected to provide important information about the
waves that may be observable with the gravitational wave detectors
currently under construction \cite{fh}. In the difficult task of
developing collision codes it has been found useful to compare
numerical relativity results with those of ``close limit''
calculations\cite{closelimit}. In these calculations the black holes
are taken to start at a small separation, and the initial separation
is used as a perturbation parameter.  The collisions of primary
interest are those with high angular momentum that end in the
formation of a rapidly rotating Kerr black hole. This is based on the
expectation that such collisions are the endpoint of binary inspiral
and that collisions with high angular momentum will produce much more
gravitational radiation than collisions leading to a nonrotating
hole. But the close limit method has so far been applied only to
collisions resulting in a nonrotating, or slowly rotating hole. There
are two reasons for this.

The first reason is a technical one. The starting point of the
dynamics of a collapse is a spatial metric $^{(3)}g_{ij}$ and an
extrinsic curvature $K_{ij}$ that solve Einstein's initial value
equations.  For the close limit perturbation method to be used for
rotating holes, these initial value solutions must be a perturbation
of a Kerr spacetime. But the prevalent method for specifying initial
data, the conformally flat prescription of Bowen and
York\cite{BowenYork}, is incompatible with the Kerr solution, for
which the standard slicing, at least, is not conformally flat. A
family of close limit perturbations of Kerr initial data, then, could
not be given in the Bowen-York formalism.

This technical difficulty is no longer a barrier.  New axisymmetric
initial data solutions representing two rotating  holes have
recently been given by Baker and Puzio\cite{bakerpuzio}, and by
us\cite{twokerrprd}.  In both types of solutions close limit sequences
of initial data can be constructed for holes starting a head-on
axisymmetric collision. More specifically, a sequence can be
constructed for different initial separations, with the close limit of
the sequence being the Kerr spatial geometry and extrinsic curvature
outside a Kerr horizon. We have chosen to
start with the close limit initial data set of
Ref.\,\cite{twokerrprd} based on the Brill-Brandt-Seidel\cite{brandtseidel} form
of the 3-metric and of the nonvanishing components of extrinsic
curvature,
\begin{equation}
  \label{genmet}
ds^2=\Phi^4\left[e^{-2q}\left(d\bar{r}^2
+\bar{r}^2d\theta^2\right)
+\bar{r}^2\sin^2\theta\,d\phi^2
\right]\ ,
\end{equation}\begin{equation}\label{extrin_r_1}
K_{\bar{r}\phi}=\bar{r}^{-2}\Phi^{-2}\widehat{H}_{E}\sin^2\theta\ ,
\hspace*{.05in}
K_{\theta\phi}=\bar{r}^{-1}\Phi^{-2}\widehat{H}_{F}\sin\theta\ ,
\end{equation}
in which $\Phi,q,\widehat{H}_{E},\widehat{H}_{F}$ are functions of
$\bar{r},\theta$.  For the appropriate choices
$\Phi_K,q_K,\widehat{H}_{EK},\widehat{H}_{FK}$, these become the
metric and extrinsic curvature for a slice of Kerr spacetime at constant
Boyer-Lindquist\cite{boyerlindquist} time. In this case the $\theta$
coordinate of (\ref{genmet}) and (\ref{extrin_r_1}) is the same as the
Boyer-Lindquist $\theta$, and the Boyer-Lindquist radial coordinate
$r$ is a rescaling of $\bar{r}$, depending on the Kerr parameters
$M,a$, as given in Ref.\,\cite{twokerrprd}.

In our approach to an initial value solution we take $q$,
$\widehat{H}_{E}$,$\widehat{H}_{F}$ to have precisely their Kerr forms
$q_K$, $\widehat{H}_{EK}$,$\widehat{H}_{FK}$. It turns out that this
guarantees that all of the initial value equations are satisfied
except one, the Hamiltonian constraint, an elliptic equation for
$\Phi$. Boundary conditions for this equation must be chosen in order
to complete the specification of a solution. One condition is
asymptotic flatness. 
If for the other condition we choose to have $\Phi\sim \kappa_{\rm
sing}/\bar{r}$ as $\bar{r}\rightarrow0$,  then we
find the Kerr solution $\Phi_K$ if $\kappa_{\rm
sing}$ is
chosen appropriately. (Otherwise we find a distorted Kerr hole.)  To
arrive at an initial value solution representing {\em two} initial
holes, we choose two point-like singularities on the $z$ axis in the
$\bar{r},\theta$, plane, and we place them at $z=\pm z_0$ (that is, at
$\bar{r}=z_0, \theta =0,\pi$). The parameter $z_0$, then, describes
the initial separation between the two holes. It is shown in
Ref.\,\cite{twokerrprd} that the $z_0\rightarrow0$ limit of the
initial solution is the Kerr geometry, outside the Kerr horizon, with
mass $M$ and spin parameter $a$ (the parameters chosen for the
functions $q_K, \widehat{H}_{EK},\widehat{H}_{FK}$).
In the limit of small $z_0$, the 3-geometry and
extrinsic curvatures are perturbations away from the Kerr forms. 
This provides the starting point for evolving perturbations
on a Kerr background.

The second reason that close limit calculations have previously been
limited to collisions forming a Schwarzschild (nonrotating) hole is
that no numerical relativity results were available of collisions
leading to Kerr formation.
This can, of course, be viewed as an opportunity to provide the only
available -- though very limited -- answers about the nature of
collisions to form Kerr holes. But our real motivation in
presenting these results is to continue the useful interaction of
perturbation studies and numerical relativity.  In particular, we hope
that these results stimulate work on fully nonlinear evolution of
initial data solutions amenable to close limit perturbation methods,
and hence to comparisons. 

Linearized evolution for perturbations of the Kerr geometry is carried
out with the Teukolsky equation\cite{Teuk73}, rather than the
Zerilli\cite{zerilli} or Regge-Wheeler\cite{reggewheeler} equation of
Schwarzschild perturbations. The introduction of the Teukolsky
equation entails two new elements in a calculation. Since the Kerr
background is not spherically symmetric, its perturbations cannot be
decomposed into spherical harmonics\cite{spinwtsph}. The dependence on
a polar angle $\theta$ cannot therefore be separated.  The linearized
numerical evolution must be carried out as a solution of a 2+1
variable ($r,\theta,t$) linear hyperbolic differential equation,
rather than a 1+1 variable ($r,t$) equation for each multipole of
spherical perturbations. (Since we are discussing the evolution of
perturbations of the Kerr spacetime, here and below the coordinates
$r,t,\theta$ are the standard, i.e.,
Boyer-Lindquist\cite{boyerlindquist} coordinates for the Kerr
spacetime.)  To solve the 2+1 partial differential equation we use an
existing and thoroughly verified 2+1 evolution
code\cite{TEUKCODE}. Though the computational requirements of
this code are much greater than for 1+1 codes, they are handled easily
by the RAM and speed of large modern workstations.  It should be noted
that the essential difficulties of numerical relativity lie in the
nonlinear equations of Einstein's theory. None of the worst
difficulties affect our linearized version of the evolution equations.

The second new feature to be faced is the way in which initial
conditions must be handled. In the Zerilli\cite{zerilli} or
Regge-Wheeler\cite{reggewheeler} formalism, evolution is carried out
for a ``wavefunction'' that is defined in terms of metric
perturbations. It is relatively straightforward to compute the initial
wavefunction and its initial time derivative from an initial value
solution for $^{(3)}g_{ij}, K_{ij}$. In the Teukolsky formalism the
quantity that is evolved with a wave equation has a somewhat different
character. This quantity originates in the
Newman-Penrose\cite{newmanpenrose} formalism that encodes information
about curvature into five complex fields $\psi_0,\ldots,\psi_4$ that
are projections of the Weyl tensor (equivalent to the Riemann
curvature tensor $R_{\alpha\beta\gamma\delta}$ in vacuum) onto a null
tetrad of complex vectors $\ell^\mu,n^\mu,m^\mu,\bar{m}^{\mu} $. Of
particular importance here, is the projection
\begin{equation}
\label{psi4def}
\psi_4\equiv  R_{\alpha\beta\gamma\delta} n^{\alpha} {\bar{m}}^\beta
              n^\gamma {\bar{m}}^\delta \; ,
\end{equation}
since the Teukolsky
formalism 
evolves  the wave function
\begin{equation}
\label{Psidef}
\Psi \equiv(r-ia\cos\theta)^4\,\psi_4 \;.
\end{equation}
It turns out that $\psi_4$ (and hence $\Psi$) is invariant to first
order under first order perturbations of the tetrad, so the
specification of initial data amounts to the computation of the
initial value of $R_{\alpha\beta\gamma\delta}$, and its initial time
derivative.  The computation of the initial $\Psi$ is carried out
using the Gau{\ss}-Codazzi equations\cite{gausscodazzi} which
relate the Riemann curvature to $^{(3)}g_{ij}, K_{ij}$ of the initial
value solution.  
Finding the initial
$\partial_t\Psi$ requires that we know the initial time derivative of
the Weyl tensor. To find this we substitute the first order vacuum
Einstein evolution equations for the 3-metric and the extrinsic
curvature into the explicit formula for $\partial_t \Psi$.  Though
straightforward in principle, the calculation of the initial $\Psi$
and of $\partial_t\Psi$ from the initial $^{(3)}g_{ij}, K_{ij}$ is
very lengthy, and subject to error.  To carry out the computation, a
{\it Maple} script was written to find $\Psi$ from the initial data
functions $\Phi,q,\widehat{H}_E,\widehat{H}_F$. The output of the
script was checked against analytic expressions known for
$a=0$\cite{CARLOS}.  For $a\neq0$ the result was checked numerically
against a very different {\it Maple} script, kindly supplied by Gaurav
Khanna, in which $\psi_4$ is computed from the linearized
perturbations on a Kerr background\cite{khannaetal}. In both
comparisons the agreement was well within the expected numerical
accuracy.

For values restricted to the region outside the event horizon of the
Kerr background, it is useful to introduce the radial variable $r^*$
defined\cite{chandrabook} in terms of the standard (Boyer-Lindquist)
radial variable $r$ by $dr^*/dr=$
$(r^2+a^2)/(r^2-2Mr+a^2)$. The $r^*$
variable agrees with $r$ in the limit of large $r$ and approaches $-\infty$
at the unperturbed horizon $r_+=M+\sqrt{M^2-a^2}$. Since the value of 
$\Psi$ diverges as $r^3$ for large $r$ we introduce 
${\cal Q}\equiv M \Psi/r^3$.

\begin{figure}
\epsfxsize=0.47\textwidth
\epsfbox{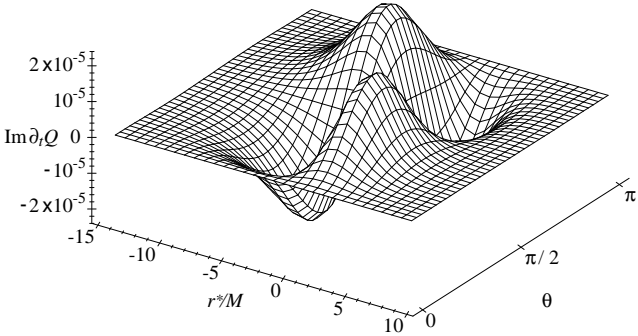}
\caption{\label{Figure-1}
Initial data for ${\rm Im} \partial_t {\cal Q}$ as a function of the
spatial coordinates $r^*$ and $\theta$. Note the antisymmetry with
respect to the equatorial plane, given by $\theta=\pi/2$.
}
\end{figure}

We have chosen to evolve initial solutions representing holes of equal
mass and spin, so that each collision is characterized by the set of
parameters $a,M,z_0$ (where $M$ and $a$ are the mass and spin
parameters of the final Kerr hole), or by the two dimensionless
parameters $a/M, z_0/M$.  Since $\Psi$ and $\partial_t\Psi$ are
complex, there are four functions of $r,\theta$ contained in them. Due
to the symmetry of the collision these functions all have symmetry
properties with respect to the $\theta=\pi/2$ equatorial plane.
Specifically, ${\rm Re}\Psi\,, {\rm Re}\partial_t{\Psi} $ are
symmetric with respect to the equatorial plane, while ${\rm Im}\Psi,
{\rm Im}\partial_t{\Psi}$ are antisymmetric. The plot of ${\rm
Im}\partial_t{\cal Q}$ as a function of $r^*$, in
Fig.\,\ref{Figure-1}, shows the antisymmetry of ${\rm
Im}\partial_t\Psi$ and the vanishing of $\Psi$ at the horizon and at
large $r$.

\begin{figure}
\epsfxsize=0.47\textwidth
\epsfbox{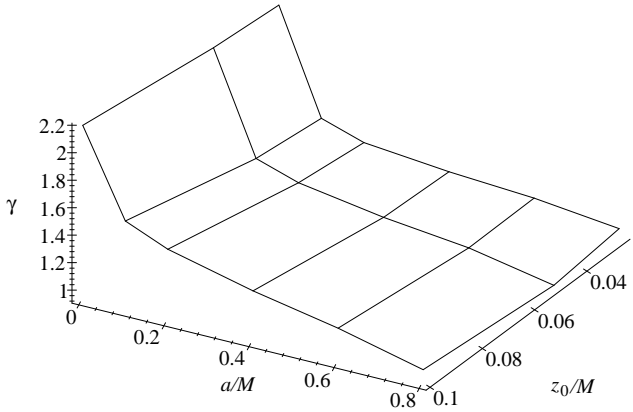}
\caption{\label{Figure-2}
  The relative deviation $\delta\equiv(\Phi-\Phi_{\rm Kerr})/\Phi$ is
  evaluated at $\bar{r}=0.5$, $\theta=0$.  For a given value of $z_o$
  we compute the relative deviation $\delta_{z_0}$, at $z_0$, and 
  the relative deviation $\delta_{2z_0}$ at $2z_0$. The exponent
  $\gamma$ is defined by
$\delta_{2z_0}=2^\gamma\delta_{z_0}$, so
  that $\gamma$ is roughly an exponent in the relationship
  $\delta\propto z_0^{\gamma}$. The figure shows that the dependence
  varies from approximately quadratic for small $a$ to approximately
  linear for large $a$.
}
\end{figure}

In the case $a=0$ the conformal factor $\Phi$ can be written in
analytic form, and it can be shown that in the $z_0\rightarrow0$
limit, the relative deviation from the Schwarzschild geometry is
quadratic in $z_0$. In the numerically computed solutions for $\Phi$,
we find that the relative deviation $\delta$ is generally linear in
$z_0$. Since this dependence must revert to a quadratic one for
$a\rightarrow0$, we infer
\begin{equation}
  \label{z0dependence}
  \delta=F_1(r,\theta;a)z_0 +F_2(r,\theta;a)z_0^2+{\cal O}(z_0^3)\ ,
\end{equation}
where $F_1$ vanishes, but 
$F_2$ does not, as  $a\rightarrow0$. A numerical confirmation
of this relationship is shown in Fig.\,\ref{Figure-2}.

\begin{figure}
\epsfxsize=0.47\textwidth
\epsfbox{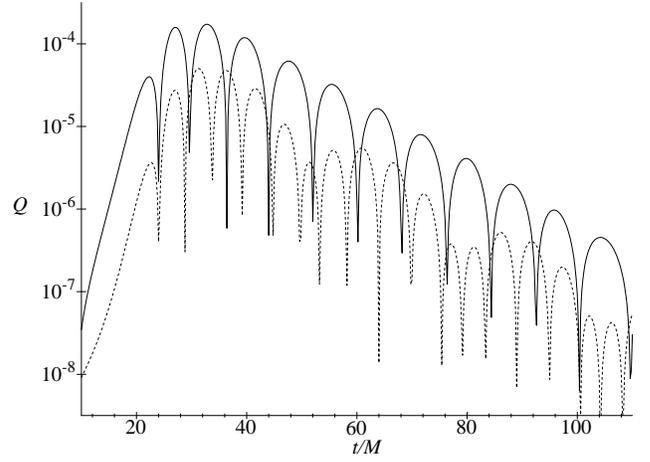}
\caption{\label{Figure-3}
Logarithmic representation of the radiation waveforms 
${\rm Re}{\cal Q}$ (solid curve) and ${\rm Im}{\cal Q}$ (dashed curve)
as a function of $t/M$ at $r^*/M=25$, 
$\theta=\pi/4$, for a collision with $a/M=0.6$, and $z_0/M=0.05$.
See text for discussion.
}
\end{figure}

The values of $\Psi$ and $\partial_t{\Psi}$ as described above are
next used as Cauchy data for the Teukolsky equation, and the equation
is solved numerically to find ${\cal Q}(r,\theta,t)$. The real and
imaginary parts of $\Psi$ correspond to the two distinct linear
polarization modes of gravitational waves, and both ${\rm Re}{\cal Q}$
(solid curve) and ${\rm Im}{\cal Q}$ (solid curve) are shown in
Fig.\,\ref{Figure-3}. A semilog plot is used to show most clearly the
nature of the oscillations and exponential damping of the waves, after
an initial transient.  Some features of Fig.\,\ref{Figure-3} have a
simple explanation: Although the Kerr background is not spherically
symmetric, perturbations at a {\em single frequency} can be decomposed
into angular functions and one can find the complex quasinormal
frequencies for each angular function. For ${\rm Re}{\cal Q}$ both the
damping (slope of the line connecting the peaks) and the oscillation
rate (spacing of zeroes) agree to within around 1\% with the frequency
of the least damped quasinormal frequency\cite{leaverthesis} $\omega_2
= (0.388 + i 0.086)/M$ for the $\ell=2$ axisymmetric mode of the
$a/M=0.6$ hole that is formed in the collision.  The explanation of
${\rm Im}{\cal Q}$ is less simple, but more interesting.  It has the
appearance of a superposition of ringing at $\omega_2$ and at
$\omega_3 = (0.618 + i 0.088)/M$. For $t/M>50$ the dashed curve in 
fact is indistinguishable from the function
\begin{eqnarray*}
&& 4.39\times10^{-4}\sin(0.388t/M+0.463)\exp(-0.086t/M)\\
 &+&8.26\times10^{-4}\sin(0.618t/M-1.951)\exp(-0.088t/M)\ .
\end{eqnarray*}
This superposition comes about because the Teukolsky
equation mixes real and imaginary parts, through a term
involving $i\,a\cos\theta
\partial_t\Psi$. The angular dependence of the imaginary part of the
initial data gives rise to small ${\rm Im}{\cal Q}$ evolution
characterized by $\omega_3$, but during evolution, the oscillations of
${\rm Re}{\cal Q}$, at $\omega_2$ are fed into ${\rm Im}{\cal Q}$ by
the Teukolsky equation. The reverse mixing of ${\rm Im}{\cal Q}$ into ${\rm
Re}{\cal Q}$ is not of importance because ${\rm Im}{\cal Q}$ is much
smaller than ${\rm Re}{\cal Q}$, and mixing is weak.

\begin{figure}
\epsfxsize=0.47\textwidth
\epsfbox{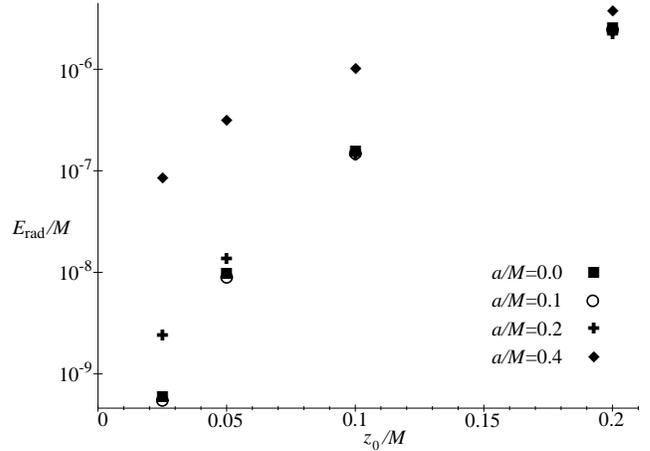}
\caption{\label{Figure-4}
The calculated total energy $E_{\rm rad}$ radiated in gravitational
waves, as a function of $z_0/M$ for different values of $a$.  A
meaningful demonstration of the influence of $a$ would require a
physical measure of separation in place of $z_0$. See text for
details.}
\end{figure}

Figure \ref{Figure-4} gives the computed total gravitational wave
energy contained in the waveforms, as a function of $z_0$ for spin
parameter values $a/M=0,0.1,0.2,0.4$. The $a=0$ problem is equivalent to
computations using Schwarzschild perturbation methods, and the results
of those methods\cite{abrahamsprice} agree with the $a=0$ points in
Fig.\,\ref{Figure-4} to better than 1.5\%.  The figure is somewhat
misleading regarding the dependence of radiated energy on angular
momentum. The variable $z_0$ is a formal measure of separation, not
directly connected with a physical separation. A physical separation,
for example the proper distance between the intersection of the axis
and the apparent horizon, depends on both $z_0$ and $a$.  Figure
\ref{Figure-4}, therefore, does not show the influence of rotation on
radiation, but it does give results for well specified initial data,
results that could be evolved with numerical relativity on
supercomputers.  We hope that comparisons with such results will be
available soon. Another application of these results will be a
comparison of predicted waveforms and energies using the slow-rotation
approximation (i.e., treating rotation as well as separation as a
perturbation), a comparison that will reveal whether slow-rotation
methods can be used for fairly rapid rotation.  We will provide such a
comparison in a more detailed presentation describing the present work.

We thank John Baker, Gaurav Khanna, and Raymond Puzio for helpful 
suggestions. This work was partially supported by NSF grant PHY-9734871.


\begin{thebibliography}{99}

\bibitem{grandchallenge} L.\ S.\ Finn, in {\it Proceedings of the 
       14th International Conference on General Relativity and Gravitation},
      Florence, 1995, edited by M. Francaviglia,
      G. Longhi, L. Lusanna, E. Sorace
      (World Scientific, Singapore, 1997), gr-qc/9603004

\bibitem{fh}\'{E}.\ \'{E}.\ Flanagan and S.\ A.\ Hughes, 
       Phys. Rev. D {\bf 57}, 4535 (1998).

\bibitem{closelimit} R.~H. Price and J.~Pullin, Phys.\ Rev.\ Lett.\ {\bf
    72}, 3297 (1994); A.~Abrahams and R.~H.~Price, Phys.\ Rev.\ D {\bf
    53}, 1963 (1996); J. Pullin, Fields Inst. Commun., {\bf 15} 117
  (1997);P.~Anninos, R.~H.~Price, J.~Pullin, E.~Seidel and W.-M.~Suen, Phys.\
Rev.\ D {\bf 52}, 4462 (1995);
J. Baker, A. Abrahams, P. Anninos, S.  Brandt, R.
  Price, J. Pullin, E. Seidel, Phys. Rev. D {\bf 55}, 829 (1997);
O. Nicasio, R. Gleiser, R. Price, and J. Pullin, Phys.\ Rev.\ D (to be
published), gr-qc/9802063

\bibitem{BowenYork}J.~Bowen and J.~W.~York, Jr., Phys. Rev. 
D {\bf 21}, 2047 (1980).

\bibitem{bakerpuzio}J. Baker and R. S. Puzio, preprint gr-qc/9802006.

\bibitem{twokerrprd} W. Krivan and R. H. Price, Phys.\ Rev.\ D {\bf
                   58}, 104003 (1998).

\bibitem{brandtseidel}S. R. Brandt and E. Seidel, Phys.\ Rev.\ D {\bf 52}, 
852 (1995); Phys.\ Rev.\ D {\bf 52}, 870 (1995); Phys.\ Rev.\ D {\bf
54}, 1403 (1995).

\bibitem{boyerlindquist}
R. H. Boyer and R. W. Lindquist, J. Math. Phys. {\bf8}, 265 (1967).

\bibitem{Teuk73}   S.\ A.\ Teukolsky, 
Astrophys.\ J. {\bf185}, 635 (1973).

\bibitem{zerilli} F.~Zerilli, Phys.\ Rev.\ Lett. {\bf 24} 737, (1971).
\bibitem{reggewheeler}T.~Regge and J.~A.~Wheeler, Phys.\
Rev. {\bf108}, 1063 (1957).

\bibitem{spinwtsph} In the Teukolsky formalism of Ref.\,\cite{Teuk73}
it is possible to separate the angular dependence of  perturbations {\em if}
perturbation functions are first decomposed into single frequency components.
Such separation is useful for some purposes, but not for time evolution
of initial data, the problem in which we are interested.

\bibitem{TEUKCODE} W.\ Krivan, P.\ Laguna, P.\ Papadopoulos, and N.\
Andersson, Phys.\ Rev.\ D {\bf 56}, 3395 (1997).

\bibitem{newmanpenrose}
E. T. Newman and R. Penrose, J. Math. Phys. {\bf3}, 566 (1962).

\bibitem{gausscodazzi} C. W. Misner, K. S. Thorne and J. A. Wheeler, {\sl
Gravitation} (Freeman, San Francisco, 1973), Sec.\,21.5.

\bibitem{CARLOS} M.\ Campanelli, W.\ Krivan, C.O.\ Lousto, Phys.\
Rev. D {\bf 58}, 024016 (1998).

\bibitem{khannaetal}
M.\ Campanelli, C.\ O.\ Lousto, J.\ Baker, G.\ Khanna,
and J.\ Pullin, Phys.\ Rev.\ D {\bf 58}, 084019 (1998).

\bibitem{chandrabook}
S. Chandrasekhar, 
{\it The mathematical theory of black holes}
(Oxford University Press, Oxford, 1983).

\bibitem{leaverthesis}Edward W. Leaver, Ph.\ D. thesis, University of 
Utah, 1985.

\bibitem{abrahamsprice}
A.M. Abrahams, R.H. Price, Phys.\ Rev.\ D {\bf 53}, 1972 (1996).

\end{thebibliography}
\end{document}